\documentclass[aps,prb,twocolumn,floatfix,superscriptaddress]{revtex4}

\usepackage{graphicx}
\usepackage{amsmath,amssymb,bm}
\usepackage{color}
\usepackage{tabls}
\usepackage[latin1]{inputenc}
\usepackage{mathptmx}
\usepackage{verbatim}

\newcommand{\expt}[1]{\langle #1 \rangle}

\newcommand{\dd}{\text{d}}
\newcommand{\ee}{\text{e}}
\newcommand{\pp}{\partial}
\newcommand{\nn}{\nonumber}

\newcommand{\LL}{\mathcal{L}}

\newcommand{\feq}{f_\text{eq}}

\newcommand{\bx}{\mathbf{x}}
\newcommand{\bp}{\mathbf{p}}
\newcommand{\bv}{\mathbf{v}}
\newcommand{\be}{\mathbf{e}}

\begin{document}

\title{Effective thermostat induced by coarse-graining of SPC water}

\author{Anders Eriksson}
\affiliation{Department of Physics, University of Gothenburg, SE-40530 Gothenburg, Sweden} 

\author{Martin Nilsson Jacobi}
\affiliation{Complex Systems Group, Department of  Energy and Environment, Chalmers University of Technology, SE-41296 Gothenburg, Sweden} 

\author{Johan Nystr\"{o}m}
\affiliation{Complex Systems Group, Department of  Energy and Environment, Chalmers University of Technology, SE-41296 Gothenburg, Sweden} 

\author{Kolbj{\o}rn Tunstr{\o}m}
\affiliation{Complex Systems Group, Department of  Energy and Environment, Chalmers University of Technology, SE-41296 Gothenburg, Sweden}

\begin{abstract}
We investigate how the transport properties of a united atoms fluid with a dissipative particle dynamics thermostat depend on the functional form and magnitude of both the conservative and the stochastic interactions. We demonstrate how the thermostat strongly affects the hydrodynamics, especially diffusion, viscosity, and local escape times. As model system we use SPC water, from which projected trajectories are used to determine the effective interactions in the united atoms model. The simulation results support our argument that the thermostat should be viewed as an integral part of the coarse-grained dynamics, rather than a tool for approaching thermal equilibrium. As our main result we show that the united atoms model with the adjusted effective interactions approximately reproduce the diffusion constant and the viscosity of the underlying detailed SPC water model.  

\end{abstract}

\maketitle

\section{Introduction}

In molecular dynamics, the addition of a thermostat is usually motivated as a representation of interactions with the surroundings, important primarily for the thermal equilibration of the system. The choice of thermostat generally depends on whether the system is characterized by constant temperature and volume (e.g. the Nos\'e--Hoover\cite{nose84a,nose84b,hoover85} thermostat) or constant temperature and pressure (e.g. the Andersen\cite{andersen80} or Parrinello--Rahman\cite{parrinello_rahman80} thermostats). Under the standard assumption that the system is sufficiently chaotic (and therefore mixing), averages over the thermodynamic equilibrium ensemble can be calculated as time averages, provided that the system is allowed to equilibrate before measuring the time average. This allows for measuring, e.g., the pressure and heat capacity of the system, and to investigate  complex phenomena such as phase transitions. Due to the fluctuation dissipation theorem, thermodynamic properties defined through the partition function are unaffected by the choice of thermostat (for further discussion, see section \ref{sec: DPD thermostat}).

In contrast to the equilibrium properties, transport processes are intimately tied to the trajectories as they depend on the auto-correlation of velocities and forces through the Green-Kubo relations~\citep{green54,kubo57}. The thermostat changes how  the trajectories approach equilibrium, and as a consequence the transport properties change as well. This is usually considered to be a problem; especially when running non-equilibrium (NEMD) simulations to measure transport properties, where the system is brought to a stationary, out of equilibrium, state by an external force field. The standard way of minimizing this effect is to make the system as large as possible, and the interaction with the thermostat as weak as possible. 

In this article we take a different view on the role of the thermostat: Rather than minimizing its effects, we view the thermostat as an integrated part of the dynamics. The thermostat is a mesoscopic representation of the exchange of energy between the coarse-grained degrees of freedom and the microscopic, rapidly fluctuating, degrees of freedom. As a result we expect the effective coarse-grained dynamics to be of Langevin type. To demonstrate this point explicitly, we study a coarse-grained model of SPC water. We use MD simulations of SPC water as our microscopic system and define the projected dynamics as the center of mass motion (united atoms) of the SPC particles. We demonstrate that it is possible to choose the effective conservative and stochastic interactions for the united atoms model, such that the equilibrium and transport properties are close to those of the projected microscopic dynamics.

The motivation for viewing the thermostat as a natural part of the system's dynamics lies in the Mori-Zwanzig theory on projection operators~\citep{mori58,zwanzig60,mori65,zwanzig02, just_etal03}. In short, the theory states that given a microscale dynamics, a lower dimensional representation can be attained through a projection of the phase space (e.g. the map from the atomic coordinates to the center of mass of the molecules), where fast degrees of freedom can either give rise to Markovian (white) noise and dissipation, or be eliminated due to averaging~\citep{zwanzig02}. Which of these two scenarios that best describe the system at hand depends on the exchange of energy (heat) between the coarse-grained degrees of freedom and the degrees lost in the coarse-graining procedure. If the energy exchange is substantial, then the motion of the coarse-grained particles will not be smooth or deterministic, but rather evolve according to a stochastic (Langevin type) differential equation. We will show that our test system is described by the latter case. Naturally, how faithfully a particular coarse-grained model represents the underlying dynamics depends strongly on the choice of projection, the mixing time of the fast degrees of freedom, and the details of the coupling between fast and slow dynamics. 

In principle, given a specific projection it is possible to derive the effective coarse-grained dynamics using the Mori-Zwanzig theory. In practice, however, the direct approach is neither computationally feasible, nor is the resulting coarse-grained system typically represented as a particle based simulation. To make the method practically useful we use the united atoms coarse-graining as an ansatz. It is important that the forces in the coarse-grained model respects the fundamental mechanical symmetries of the system, especially Galilean invariance, which implies conservation of linear and angular momentum. Therefore, we restrict ourselves to central forces that obey Newton's third law, and where the force between two particles depends only on the distance between the particles. If we additionally assume that the stochastic component of the pair-wise forces is statistically independent, we arrive at the dissipative particle dynamics (DPD) model, introduced in 1992 by \citet{hoogerbrugge_koelman92} as a simulation technique for complex hydrodynamic phenomena. In the DPD method, the thermostat is represented explicitly as dissipative and stochastic forces~\cite{hoogerbrugge_koelman92,espanol_warren95}. DPD can now be considered a standard method for mesoscopic simulation. It has been used to study, for example, complex fluids \cite{groot_warren97}, spontaneous self-assembly of amphiphilic molecules into bilayered membranes \cite{shillcock_lipowsky02}, vesicles \cite{yamamoto_etal02,yamamoto_hyodo03}, and hydrodynamics \cite{trofimov_etal02}.

We elaborate briefly on the direct consequences of choosing a Galilean invariant thermostat that faithfully characterize the local transport of linear and angular momentum. First, sufficiently close to equilibrium, the system obeys the classical result of the asymptotic $t^{-d/2}$ decay of the velocity auto-correlation ($d$ is the dimensionality of the system) \citep{ernst_etal71}. In addition, the interactions give rise to hydrodynamic modes in the fluid \citep{zwanzig_bixon70, ernst_etal71}, which lead to the Navier-Stokes equations on the macroscopic level \citep{espanol_warren95}. Galilean invariance is especially important for instance in NEMD measurements of the viscosity of complex fluids \citep{groot_warren97,soddemann_etal03}. If the equations of motion are not Galilean invariant the thermostat causes a thermodynamic screening (see, e.g. Ref.~\onlinecite{soddemann_etal03} and references therein).  A striking example of the importance of correct hydrodynamic interactions is the block co-polymer melts studied by Groot and Madden \citep{groot_madden98, groot04}, which pass through a disordered meta-stable phase before organizing themselves in a hexagonal pattern. Using DPD allowed the system to go between the two regions, whereas a similar model without hydrodynamic interactions failed to pass through the barrier. Hence, the local interactions between the coarse-grained molecules are important for the observed macroscopic state. Furthermore, the folding pathways of many proteins and peptides are sensitive to the interactions with the surrounding water, and exhibit meta-stable intermediate states separated by kinetic barriers\citep{hanggi_etal90,krivov_karplus04}.

In the standard approach to DPD the forces are usually not derived from the microscopic dynamics (for exceptions, see e.g. Ref. \onlinecite{ilpo04} and references therein). Instead, generic and simple functional forms, such as constant or linearly decreasing up to a cut-off radius, are used. Moreover, the same functional form is usually used for both conservative and stochastic forces. It should be emphasized that this choice is guided by maximizing simplicity rather than strict physical arguments (the generic repulsive nature of the conservative force is, however, consistent with the effective interactions between the center of mass of clusters of particles \cite{forrest_suter95}). To account for transport properties it has become common practice to rescale time in order to match either diffusion or viscosity. A serious and well documented pitfall of the DPD method in this version is that it cannot be tuned to match both the diffusion rate and the viscosity simultaneously \cite{espanol_warren95,coveney_espanol97,marsh_etal97,marsh_yeomans97}. In an attempt to overcome this problem, \citet{junghans_etal08} recently presented a study where a transversal component, orthogonal to the central force, was added to the DPD interactions. It was shown that  this gives freedom to tune both the viscosity and the diffusion rate independently. However, it is clear that with this ansatz the conservation of angular momentum is no longer manifest, and we should therefore expect that the hydrodynamic behavior of the system is not necessarily faithfully represented. In the current study we limit the interactions in the DPD dynamics to central forces but change the magnitude and support of the interactions. We demonstrate that we can find parameter settings where both diffusion and viscosity are consistent between the coarse-grained and the microscopic model. 

The remainder of the article is organized as follows. First, we introduce DPD as a thermostat and show how the radial dependence of the stochastic force influences the approach to equilibrium. Second, we review how to estimate the conservative force from the center of mass motion of SPC water molecules in atomistic MD simulations. Third, in the results section, we compare transport properties of the coarse-grained model to those of SPC water. Finally, we conclude with a discussion and summarize the results.

\section{The DPD thermostat}
\label{sec: DPD thermostat}

In this section we review the DPD thermostat, and illustrate how the structure of the stochastic force gives rise to Galilean invariance of the dissipative force. We also show how the structure of the noise affects the system's relaxation towards equilibrium.

In its simplest form, the equations of motion for a DPD model, with particles positioned at $\mathbf{r}_{i}$, with velocities  $\mathbf{v}_{i}$ and momenta $\mathbf{p}_{i}$, can be written as a system of Langevin equations
\begin{align}\label{eq:simple_langevin}
	\dot{ \mathbf{r} }_{i}    &= \mathbf{v}_{i}, \nonumber \\
	\dot{ \mathbf{ p } }_{i}  &= \sum_{j \neq i} \left[ \mathbf{F}_{ij}^\text{C} + \mathbf{F}_{ij}^\text{D}
										  +\mathbf{F}_{ij}^\text{S} \right],
\end{align}
where $\mathbf{F}_{ij}^\text{C}$, $\mathbf{F}_{ij}^\text{D}$ and  $\mathbf{F}_{ij}^\text{S}$ are the conservative, dissipative, and stochastic forces between particles $i$ and $j$. Both the conservative and non-conservative interactions in DPD are modeled by central forces obeying Newton's third law, ensuring that linear and angular momentum are conserved~\citep{hoogerbrugge_koelman92}. 
In DPD, the stochastic force between particles $i$ and $j$ take the form
\begin{align}\label{eq:f_stoch}
	\mathbf{F}_{ij}^\text{S} &=  \sqrt{2k_\text{B}T}\omega( r_{ij} ) \, \zeta_{ij} \, \mathbf{e}_{ij},
\end{align}
where $r_{ij}$ is the distance between particles $i$ and $j$, $\mathbf{e}_{ij} $ is the unit vector pointing from $j$ to $i$, $k_\text{B}$ is Boltzmann's constant, and $T$ is the temperature in Kelvin. The scalar function $\omega(r_{ij})$ describe how the stochastic force depends on the distance between the particles, and $\zeta_{ij}$ is interpreted as a symmetric Gaussian white noise term with mean zero and covariance
\begin{equation}\label{eq:zeta_cov}
	\expt{\zeta_{ij}(t)\zeta_{i'j'}(t')} = (\delta_{ii'}\delta_{jj'} + \delta_{ij'}\delta_{ji'})\delta(t-t'), 
\end{equation}
where $\delta_{ij}$ and $\delta(t)$ are the Kronecker and Dirac delta functions, respectively.

At thermal equilibrium the system is distributed according to the canonical ensemble
\begin{equation}\label{eq:feq}
   \feq(\bx,\bp) = Z^{-1} \ee^{-H(\bx,\bp)/k_\text{B} T},
\end{equation}
where $Z$ is the normalization term for the distribution. The equilibrium ensemble must be invariant under the equations of motion. Since $H$ is a constant of motion for Hamiltonian dynamics, the ensemble is invariant under this part of the dynamics (this is true for any ensemble where the probability of finding the system in a given micro-state depends only on the energy), but in order for the dissipative and random forces to conserve the equilibrium, we need to choose the dissipative force $F^\text{D}_i$ such that the combined contributions from the dissipative and stochastic forces cancel when acting  on the equilibrium distribution:
\begin{align}\label{eq:equil_cond}
   0 &= \LL \ee^{-H(\bx,\bp)/k_\text{B}T} \nn\\
	  &= \sum_i \nabla_{\bp_i} \cdot \left[ - \mathbf{F}^\text{D}_i + \frac{1}{2} \sum_j 2 k_\text{B}T A_{ij}(\bx) \nabla_{\bp_j}    \right]  \ee^{-H(\bx,\bp)/k_\text{B}T} \nn\\
	  &= 	\sum_i \nabla_{\bp_i} \cdot \left[ -  \mathbf{F}^\text{D}_i  - \sum_j A_{ij}(\bx) \nabla_{\bp_j} H \right]  \ee^{- H(\bx,\bp)/k_\text{B}T}
\end{align}
where $\LL$ is the Fokker-Planck operator of Eq.~(\ref{eq:simple_langevin}), and $A_{ij}$ is a $3 \times 3$ matrix given by the covariance of the total forces on particles $i$ and $j$.  The equilibrium Fokker-Planck equation (\ref{eq:equil_cond}) is commonly referred to as a fluctuation--dissipation relation. For the DPD model, this was first analyzed by \citet{espanol_warren95}. Since it must hold for all points in phase-space, the only possible solution for the dissipative force is 
\begin{equation}\label{eq:F_d_sol}
	\mathbf{F}^\text{D}_i = - \sum_j A_{ij}(\bx) \nabla_{\bp_j} H,
\end{equation}
For the DPD model the force covariance is given by
\begin{equation}\label{eq:corr_mat}
A_{ij}  = \begin{cases}
		-\omega^2(r_{ij}) \, \be_{ij} \otimes \be_{ij} & \text{when } i \ne j \\
		\sum_{k\ne i} \omega^2(r_{ik}) \, \be_{ik} \otimes \be_{ik} & \text{when }  i = j,
	\end{cases}
\end{equation}
where $\otimes$ denotes an outer product. 
Inserting Eq.~(\ref{eq:corr_mat}) into Eq.~(\ref{eq:F_d_sol}) we can write the dissipative force on a particle as a sum of pair-wise dissipative forces:
\begin{equation}\label{eq:F_d_sol2}
	\mathbf{F}^\text{D}_i = \sum_{j \ne i} \mathbf{F}^\text{D}_{ij} 
	                      = - \sum_{j \ne i} \omega^2(r_{ij}) \be_{ij} \cdot (\bv_i - \bv_j) \, \be_{ij}.
\end{equation}
Note that since $\mathbf{F}^\text{D}_i$ depends only on the velocity differences between interacting particles, it is manifestly Galilean invariant, and it is clear from the derivation above that this is a direct consequence of the covariance property of the stochastic forces [c.f. Eq.~(\ref{eq:zeta_cov})], which in turn stems from the assumption that Newton's third law applies.

A subtle point is that there is no one-to-one relation between the stochastic forces and $A$, the total force covariance matrix: Though $\mathbf{F}^\text{S}_{ij}$ appears in the Langevin equation, the dynamics depends only on $A$ (as is seen from the Fokker-Plank equation for the system), and for each choice of the force covariance matrix there are infinitely many choices of $\mathbf{F}^\text{S}_{ij}$ that obey Eq.~(\ref{eq:corr_mat}). 
For instance, the standard formulation of the DPD equations of motion contains $N(N-1)/2$ independent random variables, where $N$ is the number of particles, but it is possible to find a representation using no more than $3N$ independent random variables. In order to use this in the simulations, however, we would need to calculate the square root of the matrix $A$ in each time step. Although both representations are equally valid, the standard form of the DPD equations of motion is much more efficient. 

The dissipation--fluctuation relation [c.f. Eq.~(\ref{eq:equil_cond})] asserts that the systems thermal equilibrium is a fixed-point for the Fokker-Planck equation, or equivalently an invariant measure of Eq. (\ref{eq:simple_langevin}). In order to better understand the effect of the stochastic forces on the path to thermal equilibrium it is illuminating to study the time-evolution of the entropy of a non-equilibrium ensemble $f(\bx,\bp)$ over the phase-space. Following \citet{green52}, we can express the difference in entropy of the ensemble $f$ to the equilibrium distribution $\feq$ as 

\begin{equation}\label{eq:entropy}
  S(\infty) - S(t) = \int\dd \bx\,\dd \bp\, f(\bx,\bp) \, \log \frac{f(\bx,\bp)}{\feq(\bx,\bp)} ,
\end{equation}
which is negative for all ensemble distributions with the same support as the equilibrium distribution, and is zero if and only if the two distributions are equal. With strictly Hamiltonian dynamics, the entropy is constant in time. Intuitively, this is because the internal energy of the system needs to change in order for the ensemble to approach the equilibrium distribution, but the Hamiltonian conserves the energy. With the addition of dissipative and random forces it can be shown that entropy difference is a negative Lyapunov function on the space of ensemble distributions $f$: 
\begin{align}\label{eq:entropy-change}
  \pp_t S(t) 
  &= k_\text{B} T \int\dd \bx\,\dd \bp\, f \sum_{ij} \left( \nabla_{\bp_i} \log \frac{f}{\feq} \right)^{\text T} A_{ij} 
                                                       \left( \nabla_{\bp_j} \log \frac{f}{\feq} \right) \nn\\
  &= \frac{k_\text{B} T}{2} \int\dd \bx\,\dd \bp\, f 
  		\sum_{i\ne j} 
		\biggl[
      \omega(r_{ij}) \, \be_{ij}  \!\cdot\!
      \bigl(\nabla_{\bp_i} - \nabla_{\bp_j}\bigr)\log \frac{f}{\feq}
		 \biggr]^2,
\end{align}
which clearly is positive. In agreement with the second law of thermodynamics, the entropy continues to increase until $f = \feq$. The main point of Eq.~(\ref{eq:entropy-change}) is that it shows explicitly how the relaxation towards the equilibrium distribution depends on the structure of the noise. As we have seen above, the conservative and dissipative forces follow from the Langevin equation and the shape of the equilibrium distribution. Hence, for the purpose of capturing the transport properties of the system, only the shape of the stochastic forces remains to be specified. 

Eqs.~(\ref{eq:simple_langevin}--\ref{eq:zeta_cov}, \ref{eq:F_d_sol2}) together establishes the general form of the DPD dynamics. Both the conservative force $\mathbf{F}_{ij}^\text{C}$, or equivalently the corresponding scalar potential, and the scalar function $\omega(r)$ depend on the particular system of interest and need to be determined to obtain the correct DPD model. In practice, this is the difficult part of DPD, and also the rationale behind the often used heuristic approach for deciding the interactions. In the following section we discuss a more systematic approach for deriving the conservative interaction from a microscopic dynamics.

\section{Conservative potential reconstruction}
\label{seq:cons_pot}

Several methods exist for deriving potentials from given equilibrium radial distribution functions (RDFs) ~\cite{lyubartsev03, lyubartsev1995, ilpo04, soper96, reith_etal03, almarza_lomba03} (alternatively, direct time averaging over the fast degrees of freedom can be used, see e.g. Refs. \onlinecite{forrest_suter95,izvekov_parrinello04}). The methods for reconstructing effective potentials from the RDF rely on the result by~\citet{henderson74}, that two pairwise potentials resulting in the same RDF cannot differ by more than an additive constant. The importance of this theorem lies in the one-to-one correspondence between pairwise central force and the radial distribution function. 

In order to determine the effective potential corresponding to a given RDF, we use the inverse Monte Carlo method developed by~\citet{lyubartsev1995}. This method starts from a discretized Hamiltonian of the system,
\begin{equation}
 H = \sum_{\alpha} \Phi_{\alpha}S_{\alpha},
\end{equation}
which corresponds to using a stepwise constant potential, $\Phi_\alpha$. $S_\alpha$ denotes the number of particle pairs separated by a distance in the range $r_\alpha$ to $r_{\alpha+1}$, where $r_0 = 0$, $r_1 = dr$ and $r_\alpha = \alpha \cdot dr$. The average of $S_\alpha$ is directly connected to the radial distribution function, $g(r_{\alpha})$, by the relation
\begin{equation}
	\frac{\expt{S_\alpha}}{N(N-1)/2} = \frac{V_\alpha}{L^3} g(r_{\alpha}),
\end{equation}
where $N$ is the number of particles, $L$ the side length of the simulation box and $V_\alpha$ the volume of the spherical shell between radii $r_\alpha$ and $r_{\alpha+1}$,
\begin{equation}
	V_\alpha = \frac{4\pi}{3}\left(r_{\alpha+1}^3 - r_{\alpha}^3\right).
\end{equation}

Using a start potential, $\Phi_{\alpha}^{(0)}$, normally chosen as the potential of mean force, $\Phi_{\alpha}^{(0)} = -k_B T\ln g\left(r_{\alpha}\right)$, a Monte Carlo (MC) simulation is made, where at each time step a random particle is moved a small distance in a random direction. From the simulation, the quantity $\expt{S_\alpha}$ is measured, and the difference from the desired value, $S_{\alpha}^*$ (given directly form the wanted RDF), is calculated,
\begin{equation}
	\Delta\expt{S_\alpha} = \expt{S_\alpha} - S_\alpha^*.
\end{equation}
Also measured is the correlation in the number of particle pairs at different distances in the system, $\expt{S_{\alpha}S_{\gamma}}$, and the covariance matrix, $\expt{S_\alpha S_\gamma} - \expt{S_\alpha}\expt{S_\gamma}$. By taking these averages in the canonical ensemble, each entry in the covariance matrix can be related to the partial derivative of $\expt{S_{\alpha}}$ with respect to the potential $\Phi_{\gamma}$, through
\begin{equation}\label{eq:dsdv}
	\frac{\partial \expt{S_\alpha}}{\partial \Phi_\gamma} = -\frac{\expt{S_\alpha S_\gamma} - \expt{S_\alpha}\expt{S_\gamma}}{k_BT}.
\end{equation}
As~\citet{lyubartsev1995} suggests, this can be used in an expansion of $\expt{S_{\alpha}}$,
\begin{equation} \label{eq:S-V equation system}
	\Delta\expt{S_\alpha} = \displaystyle \sum_\gamma\frac{\partial \expt{S_\alpha}}{\partial \Phi_\gamma} \Delta \Phi_\gamma + O\left(\Delta \Phi^2\right),
\end{equation}
from which an estimate of the error in the current potential, i.e. in $\Phi_{\alpha}^{(0)}$, is given by $\Delta\Phi_{\alpha}$. Updating the potential and restarting the MC simulation with the new potential will result in a better approximation of the RDF, and after some iterations the method converges on a potential giving rise to the desired RDF. Difficulties with convergence in the method can normally be overcome by not changing the potential as much as described by $\Delta\Phi$, but instead chose the new potential as $\Phi^{(1)} = \Phi^{(0)} - \lambda\Delta\Phi$, where $\lambda \in [0, 1]$. For further discussion on the efficiency of the inverse Monte Carlo method, see e.g. Ref. \onlinecite{Murtola_etal07}.

\section{Simulations}
\label{sec: simulations}
\begin{figure} 
\centering
\includegraphics[width=8cm]{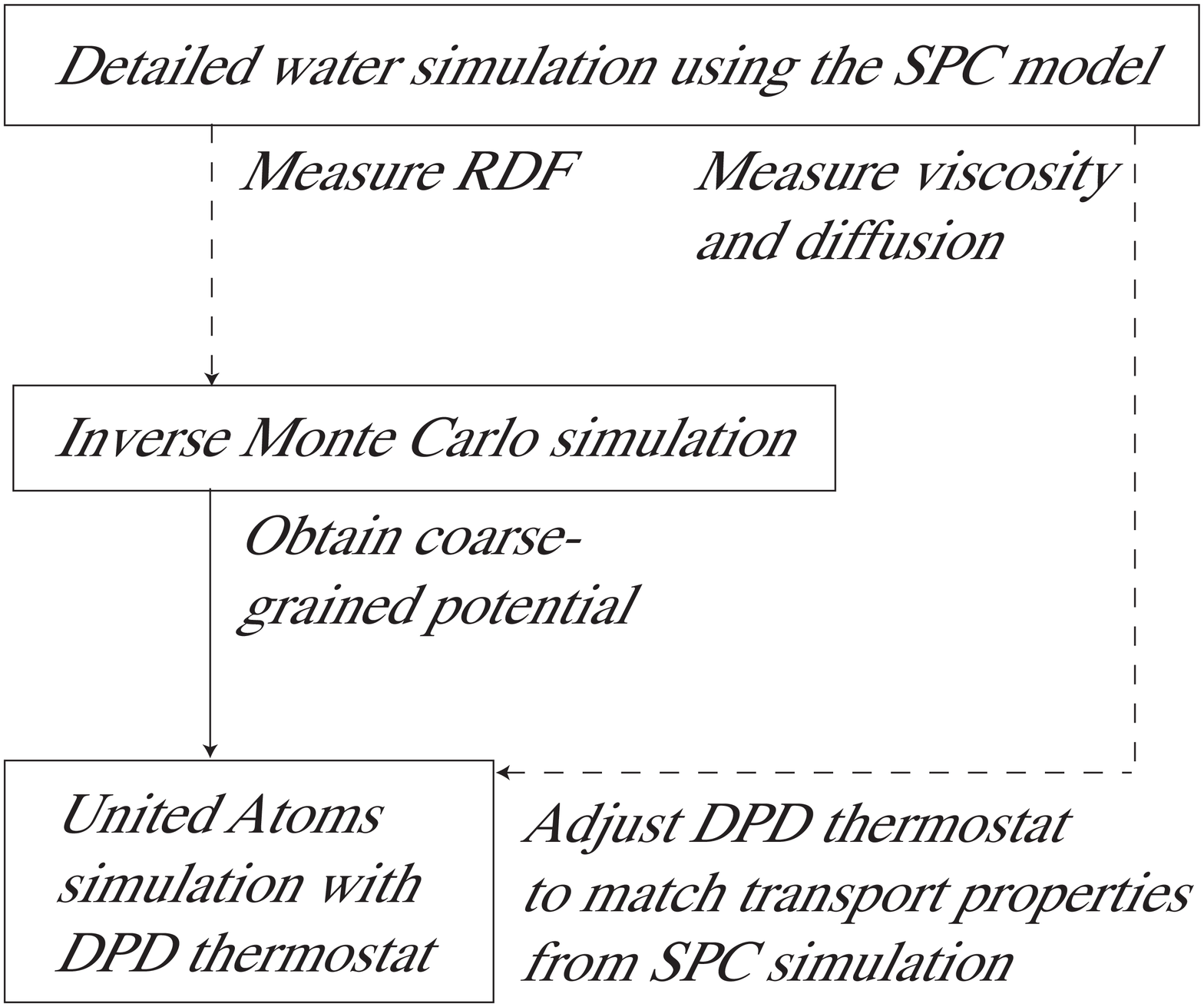}
\caption{\label{fig:outlineArticle}
An outline of the procedure used for creating effective forces for a coarse-grained representation of SPC water. The radial distribution function is measured using the center of mass positions of the SPC molecules. With the inverse Monte Carlo technique, conservative forces are then constructed, giving the correct RDF. By varying the parameters in the DPD thermostat, good agreement between the microscopic and coarse-grained dynamics can be obtained.}
\end{figure}

The procedure and technical details of our simulations are outlined below, corresponding to the schematics in Fig.~\ref{fig:outlineArticle}. From a detailed MD simulation of SPC water, coarse grained conservative forces are constructed using the inverse Monte Carlo method. By also adding a DPD thermostat we wish to demonstrate how the thermostat can be tuned to obtain good agreement between the transport properties of the coarse-grained and the microscopic levels of description. It is important to notice that the added thermostat does not affect the RDF, and therefore the conservative forces can be determined independently of the stochastic interactions. This is the rationale behind decomposing the coarse-graining procedure into two steps, as shown in Fig.~\ref{fig:outlineArticle}.

\subsection{SPC water}

Simulations using the Simple Point Charge (SPC) model of water are performed using the molecular dynamics package Gromacs. The simulation is set up with 2180 water molecules, target temperature 298 K and target pressure 1 bar (NPT ensemble with Berendsen thermostat). Electrostatic interactions are modeled using the reaction field approach for separation distances larger than $1.4$ nm. 

\subsection{United atoms}

\begin{figure}
\centering
\includegraphics[width=8cm]{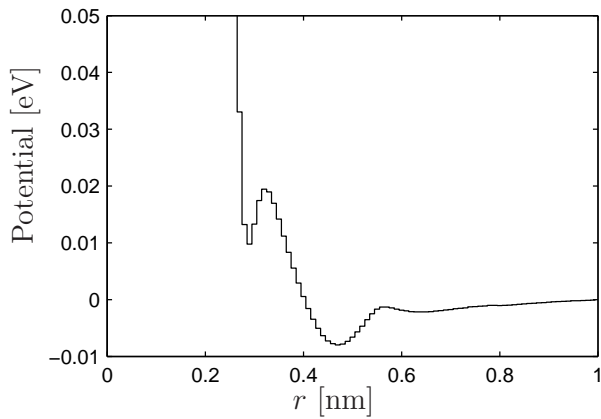}
\caption{\label{fig: coarse-grained potential} Coarse-grained potential obtained by using the inverse MC method on center of mass data from an SPC simulation. Simulation results using 2180 particles, a temperature of $298$ K, and a box with side length $4.06$ nm.}
\end{figure}

As a coarse-grained description of the same system, we use spherically symmetric point particles representing the center of mass of the SPC molecules, and assume pairwise interactions between the coarse-grained particles. This type of coarse-graining is usually referred to as United Atoms (UA). The interaction potential between the particles is obtained by applying the inverse MC method (section \ref{seq:cons_pot}) to RDF data for the center of mass of the SPC molecules. Figure~\ref{fig: coarse-grained potential} shows the coarse-grained potential as the piecewise constant approximation given by the inverse MC method. In the coarse-grained simulations, the particle density and temperature are the same as in the original SPC simulation, but we now use the NVT ensemble (Berendsen thermostat), as this makes comparison with simulations using a DPD thermostat easier. Simulations of SPC water with the NVT ensemble and box size $4.06$ nm (which was the average box size using the NPT ensemble) do not show any differences compared to the NPT simulations, so we conclude that the choice of ensemble is not important for the analysis. 

\subsection{United atoms with DPD-thermostat (UA-DPD)}

The UA-DPD simulations are performed using a standard velocity Verlet integration of the DPD equations of motion. This method gives second-order accuracy in the time-step for the conservative part, but is only first-order for the stochastic part\cite{groot_warren97}. In the case of unbounded conservative interactions for small distances, the accuracy of the integration is mainly determined by the conservative part \cite{soddemann_etal03}; therefore, the integration error in the stochastic forces is of less concern. The simulation set-up is identical to the United Atoms simulation, with the addition of dissipative and stochastic forces. We examine the simple example where $\omega(r)$ is linearly decreasing:

\begin{equation}
	\label{eq:omegaCurveLinear}
	\omega(r) = 
	\begin{cases}
		0, & 0 < r \leq r_0, \\
		\sigma  \left(1 - r/r_\text{c}\right), & r_0 < r \leq r_c, \\
		0, &  r_c < r.
	\end{cases}
\end{equation}

The parameter $\sigma$ defines the strength of the dissipative force, and $r_c$ is the cutoff radius. Note that $\omega(r)$ is also set to zero for $r\leq r_0$. The reason for this is that for the rare occasions when the inter-particle distance goes below $r_0=0.277$ nm, we do not want a stochastic force capable of pushing the particles closer together. This would result in uncontrolled magnitude of conservative forces, unless the time step is significantly lowered.

In the simulations we vary the parameters $\sigma$ and $r_c$ to examine the effect on the diffusion and viscosity values. Tuning the parameters, we try if it is possible to obtain diffusion and viscosity values consistent with those of the SPC simulation. 

\subsection{Measurements of diffusion and viscosity}

The diffusion coefficients for SPC water and the UA model without DPD thermostat are obtained directly from Gromacs. In the UA-DPD simulations, the diffusion coefficients are calculated from the mean square displacement, using the Einstein relation 
\begin{equation}
	D = \lim_{t \rightarrow \infty} \frac{\expt{[r(t)-r(0)]^2}}{6 \: t}.
\label{eq:einsteinrelation}	
\end{equation}
The viscosities for SPC and UA are measured with the NEMD method implemented in Gromacs \cite{hess02}, while the viscosity of UA-DPD is measured applying a Poiseuille flow method\cite{backer_etal05}.

\section{Results}
\label{seq:results}

\subsection{Self-diffusion and viscosity}

\begin{figure}
\centering
\includegraphics[width=8cm]{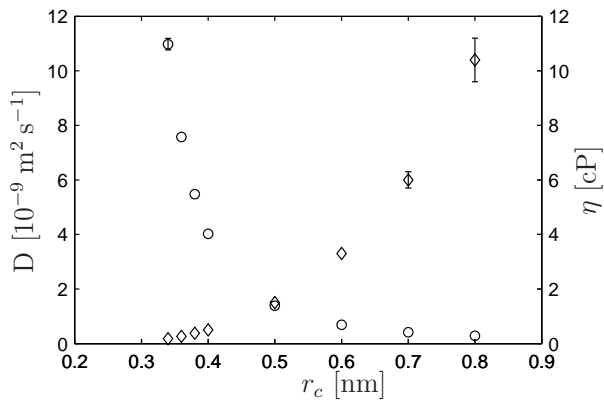}
\caption{\label{fig: diffusionAndViscosity} 
Diffusion and viscosity of UA-DPD simulations plotted as functions of the cutoff radius $r_c$. The strength of the dissipative force was identical for all simulations, $\sigma = 18.6\times10^{-17}~$ kg~m~s$^{-3/2}$. 
For the points where the error bar is not visible, the standard deviation is less than the width of the symbol.
}
\end{figure}

For the UA-DPD, we investigated the dependency of the diffusion and viscosity on the cutoff radius $r_c$. By maintaining a constant $\sigma$ for different values of $r_c$, we obtained the results shown in Figure \ref{fig: diffusionAndViscosity}. Both transport properties are strongly dependent on the value of $r_c$: While the diffusion decreases with increasing $r_c$, the viscosity increases.

The diffusion of SPC water molecules was measured to $\text{D}=4.11\times10^{-9}$~m$^2$/s. Similarly, the viscosity was measured to $\eta = 0.42$~cP. As these values represents the correct dynamics of the SPC water molecules and hence the coarse-grained representation, a coarse-grained model should be able to reproduce these numbers. The simplest approach, the UA model with a simple Berendsen thermostat, resulted in: $\text{D}=16.5\times10^{-9}$~m$^2$/s and $\eta = 0.116$~cP.  Clearly, even though this model results in the same equilibrium distribution, i.e. the same RDF, as the SPC model, it does not represent the transport properties correctly.

Adding the DPD thermostat, we tuned $\sigma$ separately for each $r_c$, such that the diffusion became approximately equal to the SPC value. The resulting $\sigma$ values and the corresponding $r_c$ values are listed in Table \ref{table: found parameters}.  Keeping $D$ constant, we examined the variation of the corresponding viscosities as a function of $r_c$ and $\sigma$. The results are summarized in Table \ref{table: results}. The viscosity is at a minimum for small values of $r_c$ and then increases moderately with increasing $r_c$. The minimum value of the UA-DPD viscosity is close to the viscosity of the coarse-grained SPC system: $0.49$~cP compared to $0.42$~cP. In particular when comparing to the UA model, it is clear that the addition of a carefully tuned DPD thermostat plays an important role in preserving transport properties. 

We would like to stress that while our choice of weight function $\omega(r)$ was an appropriate starting point, it might be too simple to capture the system in the best possible way. Our results show that the transport properties depend on the functional form of $\omega(r)$. In the light of this it is expected, rather than surprising, that the viscosity is not matched exactly. In another study\cite{eriksson_08}, we have suggested how the force covariance can be used to obtain an estimate of $\omega(r)$. A more elaborate theoretical framework based on this idea is currently in progress.

\begin{table}	
\caption{The values of $r_c$ and $\sigma$ that results in a diffusion of approximately $\mbox{D}=4.11 \times 10^{-9}$ $\text{m}^2/s$ in UA-DPD simulations. This is under the condition that the simulation is set up according to Section \ref{sec: simulations}.}
\label{table: found parameters}

\vspace{.5em}

\begin{tabular}{c@{\extracolsep{1em}}c}
	$r_c$ [nm] &  $\sigma$ [$10^{-17}$ kg m s$^{-3/2}$]\\
	\hline
	$0.36$ & $29.8$ \\
	$0.38$ & $22.3$ \\
	$0.40$ & $18.4$ \\
	$0.50$ & $10.1$ \\
	$0.60$ & $7.0$ \\
	$0.70$ & $5.4$ \\
	$0.80$ & $4.3$ \\
\end{tabular}
\end{table}

\begin{table}	
\caption{Diffusion and viscosity data for different simulations of water: SPC, UA and UA-DPD. The UA-DPD simulations are performed for different parameter values  $r_c$ and $\sigma$ of the linear weight function $\omega_L(r)$. Only the $r_c$ values are given in this table. The corresponding values of $\sigma$ are listed in Table \ref{table: found parameters}. Data for real water is included as a comparison. All simulation measurements are done at $298$ K, the water data is valid for $298.2$ K.}
\label{table: results}

\vspace{.5em}

\begin{tabular}{l@{\extracolsep{1em}}ccc}
	Method	&	$r_c$ [nm]	&	D [$10^{-9}$ $\text{m}^2 \text{s}^{-1}$]	&	$\eta$ [cP]	\\
	\hline
	SPC   	&	-	&	$4.11$	&	$0.42 \pm 0.004$ \\
	\hline
	UA	&	-	&	$16.5	$	&	$0.116 \pm 0.007$ \\
	\hline
	UA-DPD	&	$0.36$ 	& $4.16 \pm 0.07$	&	$0.51 \pm 0.03$ \\
	UA-DPD	&	$0.38$  & $4.10 \pm 0.10$	&	$0.49 \pm 0.02$ \\
	UA-DPD	&	$0.40$ 	& $4.10 \pm 0.05$	&	$0.49 \pm 0.02$ \\
	UA-DPD	&	$0.50$	& $4.07 \pm 0.07$	&	$0.52 \pm 0.03$ \\
	UA-DPD	&	$0.60$	& $4.10 \pm 0.06$	&	$0.52 \pm 0.03$ \\
	UA-DPD	&	$0.70$ 	& $4.10 \pm 0.05$	&	$0.55 \pm 0.03$ \\
	UA-DPD	&	$0.80$	& $4.08 \pm 0.07$	&	$0.60 \pm 0.04$ \\
	\hline
	Water	&	-	&	$2.27^[$\cite{eisenberg69}$^]$	&	$0.8909^[$\cite{iapws}$^]$ 
\end{tabular}
\end{table}

\subsection{Escape time distribution}

\begin{figure*}
\centering
\begin{tabular}{c}
\raisebox{5cm}{\textbf{A}}\includegraphics[width=8cm]{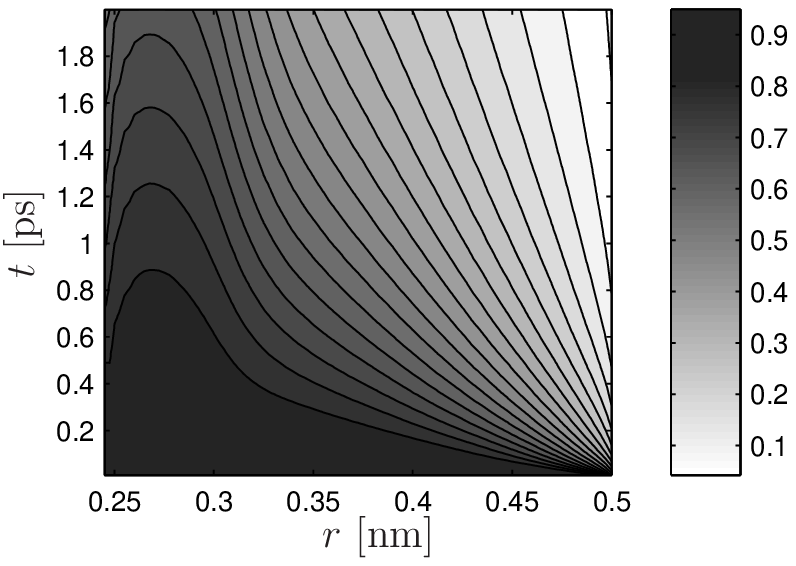} \hspace{2em}
\raisebox{5cm}{\textbf{B}}\includegraphics[width=8cm]{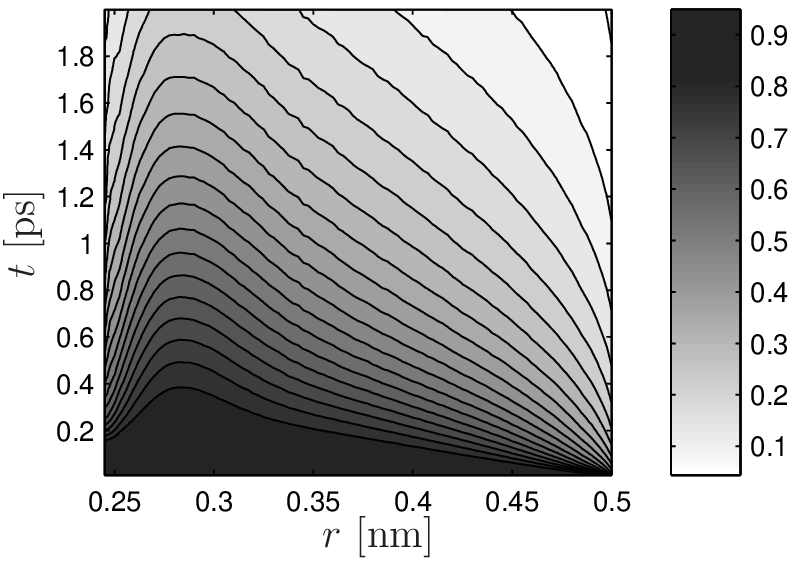} \\
\raisebox{5cm}{\textbf{C}}\includegraphics[width=8cm]{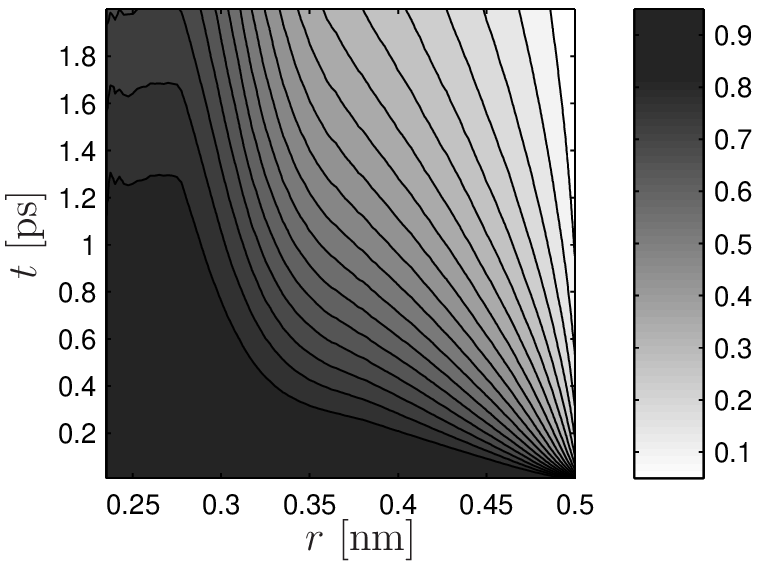}
\end{tabular}

\caption{\label{fig:escape_time}
Probability for a pair of particles to remain within a distance of $0.5$~nm from each other during a time interval $[0,t]$, given the distance $r$ between the particles at time $0$. The different figures show the escape time for (A) SPC water, (B) UA simulation with conservative force only, and (C) UA-DPD using $\omega(r)$ from Eq.~(\ref{eq:omegaCurveLinear}) with $r_c=0.38$ nm and $\sigma = 22.3 \times 10^{-17}$ kg m s$^{-3/2}$.
}

\end{figure*}

In addition to determining the diffusion coefficient and the viscosity, we have also measured the escape probabilities for pairs of particles, that is, how large the probability is for two particles to separate beyond a given distance in a given time interval, given the initial separation. This gives a more detailed view of the local dynamics than the other transport properties. 
The escape probabilities for SPC water, united atoms with the Berendsen thermostat, and the UA-DPD using $\omega(r)$ from Eq.~(\ref{eq:omegaCurveLinear}) are shown as level plots in Fig.~\ref{fig:escape_time}. The different levels in these figures represent the probability that two particles, originally separated by the distance $r$, remain within a separation distance of $0.5$~nm at a later time, $t$, displayed on the $y$-axis. 

Generally, the escape time is decreasing with increasing initial separation $r$. Note especially that the average escape time is much larger in the region $r \lesssim 0.3$~nm than for $r > 0.3$~nm. This is consistent with the well-known caging effect. Particles repeatedly bounce against their closest neighbors in the fluid, and for a pair of nearby particles to separate beyond the typical neighbor distance (about $0.28$~nm for SPC water) requires either a concerted motion of several particles, or a very high kinetic energy of the particles. Note that this barrier is largely missing from the united atoms system with the Berendsen thermostat; this is the main reason why the diffusion rate is too high compared to the SPC water. In contrast, the UA-DPD thermostat fitted to the diffusion rate and viscosity (Fig.~\ref{fig:escape_time}C) exhibit escape times similar to those of the SPC model. Hence, matching diffusion and viscosity seems to confer agreement with respect to more general transport properties.

\section{Summary and discussion}

We have applied a coarse-graining scheme to the SPC water model. The effective dynamics of the center of mass of the water molecules is Langevin-like with a conservative (drift) component. The conservative part of the dynamics is represented by a pairwise potential that can be determined from the radial distribution function. Stochastic and dissipative forces, motivated by the Mori-Zwanzig projection operator formalism, are also included as a DPD thermostat. We have demonstrated that the radial dependence of the stochastic and dissipative forces strongly affect the transport properties. The main conclusion is that the diffusion rate and the viscosity of the original system can be approximately matched by the coarse-grained model. This was obtained by tuning the magnitude and functional form of the random and dissipative forces.

We argue that in meso- and microscale simulations of coarse-grained Hamiltonian systems, the stochastic and dissipative components of the interactions should normally be considered as an integrated part of the effective dynamics. For coarse-grained models with only pairwise interactions, a dissipative particle dynamics ansatz is well suited for representing the non-conservative part of the dynamics, since it is the most general pairwise interaction that respects the local conservation or transport of linear and angular momentum, i.e. hydrodynamic behavior. The DPD interactions depend on the inter-particle distance. We argue that this functional dependence should the tuned so that the transport properties are consistent with the microscopic dynamics. More generally we show explicitly in Eq.~\ref{eq:entropy-change} how the relaxation towards equilibrium depends on the structure of the thermostat. Finally, we point out that all practical molecular dynamics methods, not only united atoms models, are ultimately coarse-grained from some more fundamental level, either from quantum mechanics or by removing fast degrees of freedom (e.g. vibration modes in covalent bonds) from a classical mechanics model. The interaction with these fast degrees of freedom plays an active role when the system equilibrate. The approach used in this paper can therefore possibly also be applied to atomistic molecular dynamics simulations.

\noindent {\bf Acknowledgements:} 
We thank Aldo Rampioni at the University of Groningen, Netherlands. This work was funded (in part) by the Programmable Artificial Cell Evolution project (EU integrated project FP6-IST-FET PACE), by EMBIO (a European Project in the EU FP6 NEST Initiative), and by the Research Council of Norway. 

\vfill

\bibliography{article}

\end{document}